# MOSFET GIDL Currentby Designing a New BTBT Model Using De-Casteljau's Algorithm

ArneshSen and Jayoti Das, Jadavpur University, Kolkata - India

*Abstract*—The Band-to-Band tunneling probability strongly depends on the shape of the potential barrier. However, parabolic approximation of this shape is well acceptable but unfortunately significant amount of error is unavoidable by using quadratic polynomial in calculation of tunneling probability. De-Casteljau's algorithm, followed by Bezier Curve can be modeled to any arbitrary shape using its Geometry Invariance Property and End points geometric property. Using this algorithm a new Band-to-Band tunneling model is designed and verified by establishing an analytic expression of Gate Induced Drain Leakage current in MOSFET.

*Index Terms*—Band-to-Band tunneling (BTBT), graded junction, parabolic approximation, gate-induced-drain-leakage (GIDL).[1]

## I. Introduction

SIGNIFICANT Amount of leakage current, called Gate-Induced-Drain-Leakage (GIDL) is observed where the gate overlaps the drain junction as device scaling continues towards deep sub-micrometer region. Band-to-Band-Tunneling (BTBT) mechanism is deeply dominant here to produce GIDL current. Again this BTBT mechanism is strongly relying on the shape of the potential barrier. This barrier shape was approximated earlier as triangular and recently as parabolic by the previous researchers. To make our problem more crystalline it is highly necessary to emphasize the reason why the triangular potential barrier is re-approximated as parabolic potential barrier. The concept of triangular potential barrier was originated in the case of abrupt junction approximation where diffused impurity profile is very steep. More closer and accurate analysis reveals that impurity profile is actually spread out into the sample which will result graded junction and as well as blunt band edges. The scope of converting any kind of approximation into accuracy is counted as a step towards the success. This typical mentality forces the researchers to generalize the impurity doping profile and the concept of spreading of impurity concentration throughout the sample is dominant. Now the appearance of the potential barrier is more likely to be parabolic rather than triangular. As the edges become blunter the parabolic shape of the barrier is more acceptable. According to J-Chen's research [1] this sub breakdown leakage current is influenced by impact ionization and electron tunneling current from the gate. To predict the mathematical expression of the tunneling current parabolic potential barrier were predicted. Moreover, Kuo-Feng and Ching-Yuan Wu also developed a tunneling current model [2] and compared it with Endoh's Model [3] and experimented data and it resulted a better match. The model, developed by Ja-Hao Chen, Shyh-Chyi Wong and Yeong-Her Wang is handy [4] enough but 8% deviation was seen when it is compared to Chen's Model [1]. At last the hard work of Xiaoshi Jin, Xi Liu and Jong-Ho Lee at last gave birth to an almost error free model [5] by calculating the net magnitude of involved electric field but that model includes rigorous mathematical calculation which was too hectic to frequent application. Another well acceptable electric field expression using work function engineering calculated by Farkhanda Ana and this expression [6] is applied in Kane's Tunneling current model [7] to calculate GIDL current. All these previous efforts have been focused on either electric field distribution or work function engineering and all the results were verified by calculating an analytic expression of GIDL current and most of the cases Kane's Tunneling Current model [8] was used. In addition to that it should be noted all of the previous researchers used the tunneling probability expression from the result of WKB approximation [9][10] using parabolic potential barrier. In fact to calculate analytical model of electrical characteristic of TFET [11] parabolic potential barrier approximation was used by Kumar and his team as well as Marie Garcia Bardon and her team tried the same approximation for designing pseudo 2D TFET [12]. Well, to sum up, it can be seen that where there was an occurrence of band to band tunneling, either parabolic or quadratic polynomial approach is applied in potential barrier equation.

  The main objective of this paper is to approximate the shape of the potential barrier in a totally different way and to verify the validity of the proposed shape by calculating an analytic expression of GIDL current and comparing the current values with existed results. De-Casteljau's algorithm to calculate the polynomials in Bezier Curves form is used as a novel weapon to solve this problem. Cubic Bezier Curves were successfully used by M. Sarfraz and his team to capture outlines of 2D shape [13] and Conic Approximation of planner curves by Y.J. Ahn [14] actually injected the virus to approximate the shape of the potential barrier using that same Bezier Curve should definitely result a more accurate expression. To hunt that expression Bezier curves are set to model the shape of the above mentioned parabolic potential barrier. After calculating the potential barrier expression WKB method is used to calculate the electron Tunneling Probability and from that ultimate expression of GIDL current is expressed. In Section-II GIDL mechanism is explained in details and Section-III contains the details of model development process. Section-IV contains the final GIDL current expression and the result is verified in Section-V.

---

[1]Manuscript submission date . This work was supported by Dept. of Physics, Jadavpur University, Kolkata – India.
ArneshSen and Prof. Jayoti Das are with Dept. of Physics, Jadavpur University, Kolkata – India
 (email: senarnesh.elec@gmail.com ; jayoti.das@gmail.com )

## II. Physical Phenomenon of GIDL

In the context of micromini when conventional lateral MOSFET is significantly scaled down significant amount of leakage current is observed due to GIDL[15] and Body Leakage[16] where the gate overlaps the drain junction. Figure-1 shows the depletion regions formed due to external voltage $V_{DG}$ (Negative Gate voltage and Positive Drain Voltage). A close look of Figure 1(b) reveals that the depletion region is extended in drain region at gate-drain overlapped portion due to positive $V_{DG}$. This additional depletion region will introduce the rising of bang bending for $V_{DG}>0$. The generation of GIDL current is completely depends upon the drain doping concentration. The doping concentration should not be too low to tunnel and too high to reach band-bending drops below the Si band gap value $E_G$. Typically for BTBT mechanism moderate ($\sim 10^{18}$) drain doping concentration should be maintained to produce GIDL. The generation of electron-hole pair is dependent on critical field strength within the extended depletion region. Covalent bond electrons will be torn out leaving free holes behind and will accelerate towards positive plate of $V_{DG}$ through n$^+$ drain. Figure-1(a) depicts that there will not be any occurrence of band bending as Drain and Gate is not overlapped. In Figure-1(b) a Band Bending is occurred due to negative gate and generation of leakage current is self-explanatory.

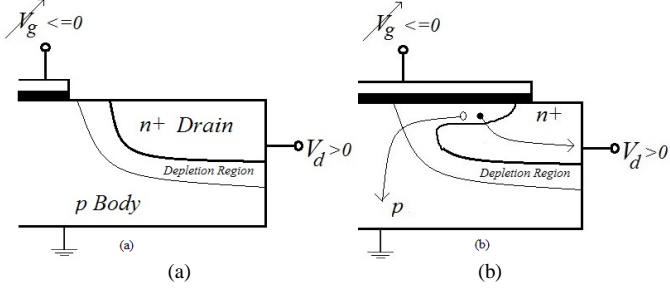

(a)  (b)

Figure-1: Cross sectional view of (a) Non-Overlapped and (b) Overlapped Gate-Drain Junction

## III. Model Developing for potential barrier

### A. For intermediate Control Point is fixed

We have the equation of the potential barrier with an intermediate control point and to make our model more convenient to the shape-variation of the potential barrier we put a constraint that the intermediate control point can move only horizontal direction by setting $y_2$ to zero. As the vertical movement of this control point results most of the non-realistic arbitrary shapes. For the sake of simplicity it is assumed that the horizontal movement of $x_2$ is restricted at origin. Figure-2(a) explains the conventional band diagram for BTBT mechanism and Figure-2(b) reveals how the conventional band diagram can be fitted in our proposed model. An incoming electron with energy $E_e$ can tunnel through this potential barrier and the classical tunneling points are ($-x_1$,0) and origin (0,0) , where $x_1$ is the tunneling width.. The conventional parameters of the Bezier Curve equation should be perfectly compatible with the desired Potential Barrier Equation as tabulated (See Table – I). By following the standard steps of Bezier Curve Parametric equation to get desired equation (See Appendix – I) and from TABLE-1 we can get the approximate equation (Equation No ) of the desired potential barrier for our model. Now using very popular approximation of WKB we can estimate the desired tunneling probability. For detail calculation see Appendix 2.

TABLE – I
COMPATIBILITY TABLE OF PROPOSED BEZIER CURVE MODEL PARAMETERS

| Parameters | Symbols for Conventional Bezier Curve | Symbols for Desired Potential Barrier Model | Descriptions of the Symbols of Column 3 for this Table. (See Figure) |
|---|---|---|---|
| Control Points | $x_3$ | $x_1 + x_3 = w_d$ | DEPLETION WIDTH OF THE JUNCTION |
| | $x_1$ | | |
| | $x_2$ | $x_2$ | PARAMETER, PROPORTIONAL TO DOPING PROFIL CONCENTRATION. |
| | $y_3$ | $V_b$ | PARAMETER, PROPORTIONAL TO BAND BENDING POTENTIAL |
| | $y_2$ | ZERO | SET IN ORIGIN |
| | $y_1$ | ZERO | SET IN ORIGIN |

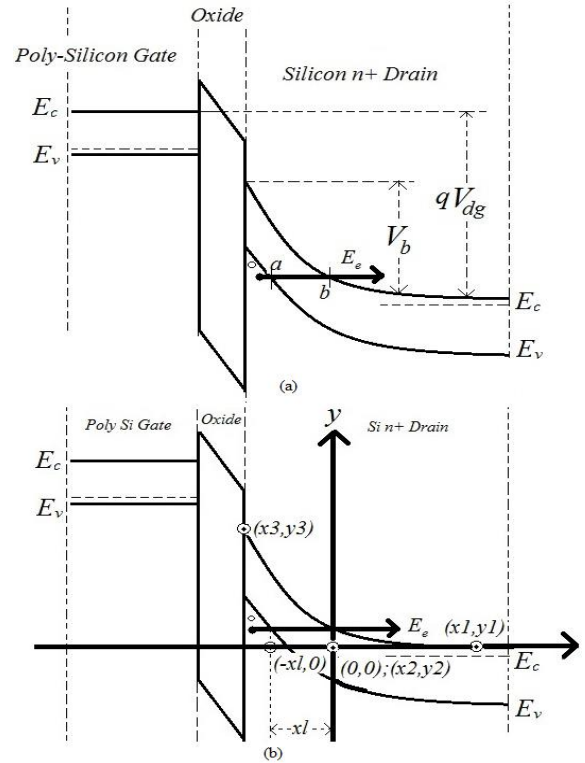

Figure 2: (a) BTBT Process (b) BTBT process Model Setup using Bezier Curve.

Using the conventional steps to construct any specific equation from the control points of Bezier curve the equation of required function have been found as follows:(For details steps see Appendix 1.)

$$y(x) = \frac{(x-x_1)^2 y_3}{4(x_1-x_2)\{w_d-(x_1+x_2)\}} \quad (1)$$

where $w_d = x_3 + x_1$.

Now for simplicity the equation is normalized at first and then $x_2$ and $x_1$ is set to zero and ½ respectively. Thus the normalized equation becomes,

$$y(x_n) = \left(x_n - \frac{1}{2}\right)^2 y_3 \quad (2)$$

All the parameters with suffix 'n' are the normalized parameters. All normalization is done with respect to depletion width $W_d$(More simply we can say to normalized we have to divide both numerator and denominator with $W_d$). Interestingly this equation also came out as quadratic equation but not exactly same as previous equations which were used in previous models [12][13]. With the help of Table-1 we can replace the normalized control points by physical factors responsible for tunneling. By replacing we can have our potential barrier equation which is as follows:

$$V(x_n) = \left(x_n - \frac{1}{2}\right)^2 AV_b \quad (3)$$

Obviously vertical axis denotes the potential and A is the proportionality constant which value is set to 400 for model validation. Now, using most familiar steps of WKB approximation[16] we can derive the tunneling probability using our potential barrier equation within the limits of the classical tunneling points '*a*' and '*b*'(See Figure-2a) and incoming electron energy is set to $E_e$ (See Figure − 2b). According to WKB approximation tunneling probability can be written as:

$$T(E_e, x_l) = \exp\left[-\frac{2m_o}{\hbar^2}\beta\left\{\int_a^b \{V(x_n) - E_e\}^{1/2} dx\right\}\right] \quad (4)$$

Where β is one of the model parameter. Now putting the value of $V(x_n)$ from equation (3) in equation (4) and by setting classical tunneling points from $-x_{nl}$ to 0 (Figure-2b) we get,

$$T(E_e, x_{nl}) = \exp[-\beta\left(\frac{2m_0}{\hbar^2}\right)^{1/2} \int_{-x_{nl}}^{0} \left\{AV_b\left(x_n - \frac{1}{2}\right)^2 - E_e\right\}^{\frac{1}{2}} dx] \quad (5)$$

Where $x_{nl}$ is the normalized value of $x_l$.
By simplifying equation 5 we get, (See Appendix-2)

$$T(E_e, x_{nl}) = \exp\{-F(T_1 - T_2)\} \quad (6)$$

Where,

$$F = \left(\frac{2m_0}{\hbar^2}\right)^{1/2} \beta \quad (6.1)$$

$$T_1 = \left(\frac{AV_b}{2}\right)^{1/2} \left[\left(x_{nl} + \frac{1}{2}\right)\left\{\left(x_{nl} + \frac{1}{2}\right)^2 - \frac{E_e}{AV_b}\right\}^{\frac{1}{2}} - \frac{1}{2}\left(\frac{1}{4} - \frac{E_e}{AV_b}\right)^{\frac{1}{2}}\right]$$

$$T_2 = \frac{E_e}{2(AV_b)^{\frac{1}{2}}} \ln\left\{\frac{\frac{1}{2}(AV_b)^{\frac{1}{2}} - \left(\frac{AV_b}{4} - E_e\right)^{\frac{1}{2}}}{\left(x_{nl} + \frac{1}{2}\right)(AV_b)^{1/2} - \sqrt{\left(x_{nl} + \frac{1}{2}\right)^2 AV_b - E_e}}\right\}$$

### B. For variable intermediate control point

Equation 6 is our tunneling probability with xn2=0. Now we should investigate what happens when xn2 is not equal to 0 but varies in negative x axis. $x_{n2}$ is one of the normalized control points of the potential barrier and according to our model if we free this control point the shape of the potential barrier is modulated. So in mathematics if we simply use normalized version of equation 1 instead of equation 2 our purpose is solved. The normalized version of equation 1 by setting $x_{n1}=1/2$ ($x_{n1}$ is normalized value of $x_1$) we get,

$$y(x_n) = \frac{\left(x_n - \frac{1}{2}\right)^2 y_3}{4\left(\frac{1}{2} - x_{n2}\right)^2} \quad (7)$$

And corresponding modified potential barrier equation is (See Table-1),

$$V(x_n) = \frac{\left(x_n - \frac{1}{2}\right)^2 AV_b}{4\left(\frac{1}{2} - x_{n2}\right)^2} \quad (8)$$

Now similarly to find tunneling probability with $x_{n2}$ we have to put the value of $V(x_n)$ from equation (8) to equation (4). Now the question arises that what should be the modified classical points if we include $x_{n2}$ in potential barrier equation. To get answer we have to concentrate the following figure where the change of shape of potential barrier is shown with respect to $x_2$.

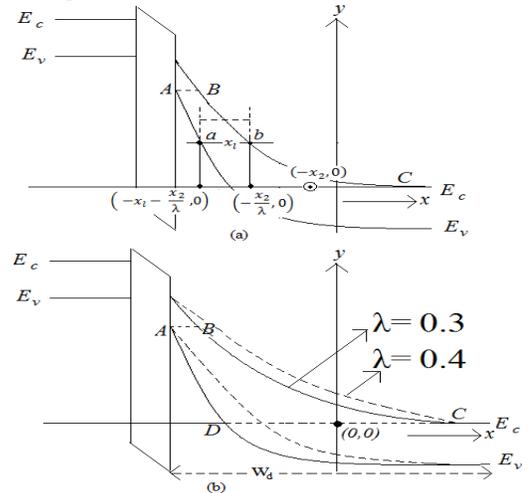

Figure-3: Variation of (a) Classical Tunneling Points and (b) Potential Barrier with λ

If we take a closer look on Figure-3 we can easily differentiate among the maximum possible tunneling region ABCDA which is inversely proportional to a new parameter λ. Now from Figure-3 it is clear that depletion width $W_d$ is constant over every ABCD region but the position of $x_2$ is varied. The normalized value of $x_2$ is $x_2/W_d=x_{n2}$. The coordinates of the modified classical tunneling points are
$(-x_a,0)$ and $(-x_b,0)$. Thus $x_a$ and $x_b$ should be λ dependent parameters and for this model they are related as follows.

$$x_b = \frac{x_{n2}}{\lambda} = \frac{x_2}{W_d \lambda} \quad or, \quad x_2 = \lambda w_d x_b$$

With the help of equation one can easily relate how λ influence the ABCD region under the supervision of the natures of Bezier Curves.

Now we have our new limits of the classical points. From equation (4) and (8) we can rewrite the Tunneling Probability expression with new limits as follows:

$$T(E,x) = \exp[-F \int_{-x_l-\frac{x_2}{\lambda}}^{-\frac{x_2}{\lambda}} \{V(x_n) - E\}^{\frac{1}{2}} dx_n] \qquad (9)$$

Now after putting the value of v(x) from equation (8) to equation (9) we can solve the integration (Appendix-2). After solving we get

$$T(E_e, x_{nl}, x_{n2}) = \exp\{-F(T_{11} - T_{22})\} \qquad (10)$$

Where,

$$T_{11} = \frac{AV_b}{(2-4x_{n2})}[\left(x_{nl} + \frac{x_{nl}}{\lambda} + \frac{1}{2}\right)\sqrt{\left(x_{nl} + \frac{x_{n2}}{\lambda} + \frac{1}{2}\right)^2 - \frac{4E\left(\frac{1}{2} - x_{nl}\right)^2}{AV_b}}$$

$$-\left(\frac{x_{n2}}{\lambda} + \frac{1}{2}\right)\sqrt{\left(\frac{x_{n2}}{\lambda} + \frac{1}{2}\right)^2 - \frac{4E\left(\frac{1}{2} - x_{nl}\right)^2}{AV_b}}$$

$$T_{22} = \frac{E\left(\frac{1}{2}+x_{n2}\right)}{\sqrt{AV_b}} \ln \frac{\left(\frac{1}{2}+\frac{x_{n2}}{\lambda}\right)\sqrt{AV_b} - \sqrt{\left(\frac{1}{2}+x_{n2}\right)^2 \frac{AV_b}{4} - 4E\left(\frac{1}{2}-x_{n2}\right)^2}}{\left(x_{nl}+\frac{x_{n2}}{\lambda}+\frac{1}{2}\right)\sqrt{AV_b} - \sqrt{\left(x_{nl}+\frac{x_{n2}}{\lambda}+\frac{1}{2}\right)^2 AV_b - 4E\left(\frac{1}{2}-x_{n2}\right)^2}}$$

And the value of F is given in equation (6.1). Now our equation of tunneling probability (T) with and without $x_2$ is ready. We can verify our model by varying incoming electron energy $E_e$ and Normalized Tunneling width with Tunneling Probability of equation 6 with different bending potential. The three dimensional plot of Figure-4 will justify our model validity. From the band diagram it is clear that $V_{dg}$ (gate to source voltage) has direct influence on bending potential and that bending potential is one of the most important parameter for tunneling amplification (Figure-5). To illustrate the influence of $V_{dg}$ over $V_b$ we used the relation derived by the previous researchers[6] as follows:
$V_b =$

$$V_{dg} - V_{fb} + \frac{qN_d T_{ox}^2 \epsilon_{si}}{\epsilon_{ox}^2} - \sqrt{\left(\frac{V_{dg}-V_{fb}+qN_0 T_{ox}^2 \epsilon_{si}}{\epsilon_{ox}^2}\right)^2 - \left(V_{dg}-V_{fb}\right)^2} \quad (11)$$

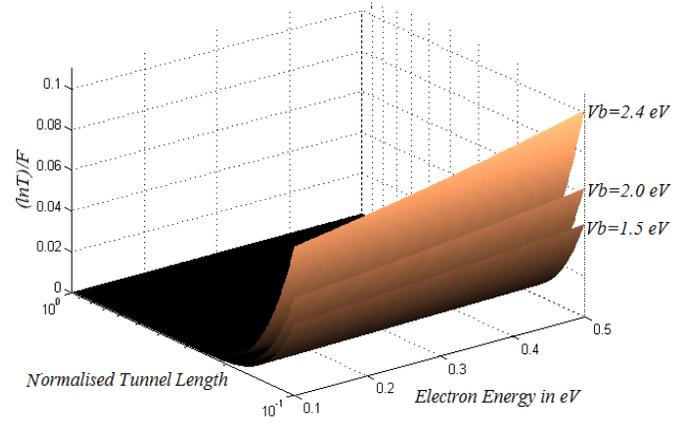

Figure-4: Variation of Tunneling Probability with tunnel length and electron energy.

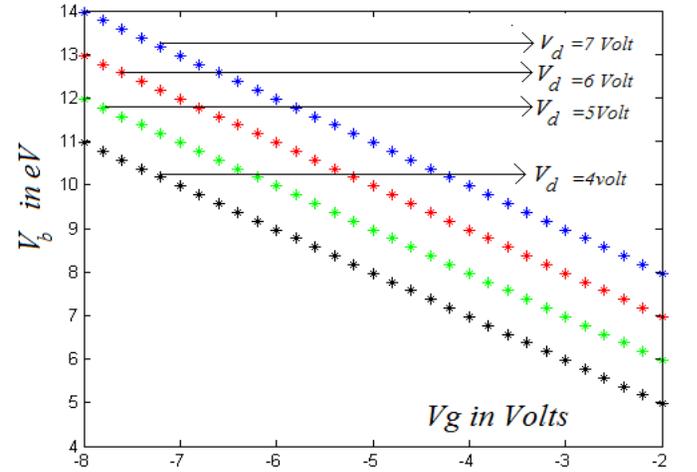

Figure-5: Influence gate voltage over bending potential

.Now let us focus on our Tunneling Probability Expression which includes $x_{n2}$ (Equation-10). From Figure we already noticed that the structure of tunneling region which means how fast amount of tunneling probability increases with the incoming electron energy is completely depends upon λ. Thus this λ should have a certain physical significance. The existence of graded junction is well familiar and it is also common that the bluntness of the barrier edges of the junction is dependent on doping concentration [17]. The derivations become complicated when we deals with this kind of graded index profile. Exactly at this point the second unique feature of our model is exposed. If we compare the graded junction bluntness variation with Figure a interesting match is noticed and from this point of view we can predict the new parameter λ may be compared to the physical variation of doping concentration. For instance from Figure we can say that with λ=0.3 the transmission probability varies more rapidly than the transmission probability with λ=0.4 as with the increasing amount of incoming electron energy the tunnel width decreases and this decreasing rate is dependent upon λ. If we vary Transmission Probability with incoming electron energy even in small amount (Physically in tunneling window the scope of incoming energy variation is very small also) from equation 10 we get Figure. The plot of figure is the best evidence of our prediction where the curves of different set based upon λ cross across each other reveals clearly the rapidness of transmission probability. Well, the exact calibration of λ with impurity doping profile is apparently beyond the scope of this paper but to expose the capability of our predicted model it is evident that an impurity profile based model with variable grade constant can be planned using equation (10).

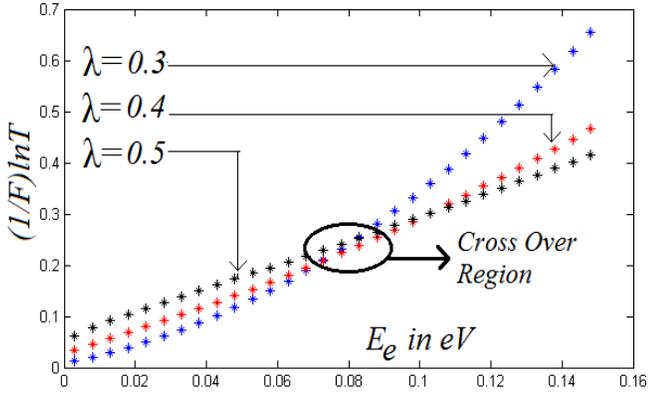

Figure-6: Effect of λ over Tunneling Probability

## IV. GIDL CURRENT EXPRESSION

The resultant current density can be related to transmission probability as [18]

$$I_{GIDL} = A_d J(E_e) = A_d \alpha \frac{q m_0}{2\pi^2 \hbar^3} T(E_e, V_b) \int_{E_e}^{\infty} \{f_v(E) - F_c(E)\} dE \quad (12)$$

Where $m_0$ is the rest mass of electron, $\hbar$ is reduced plank constant and $A_d$ is the effective area. $f_v$ and $f_c$ are probability distribution function of valence and conduction band respectively. By assuming valence band full of electrons and conduction band with no electron we can put $f_v=1$ and $f_c=0$. The value of $T(E_e, V_b)$ is taken from equation (9) and α is another model parameter. Using equation (6), (11) and (12) we can express $I_{GIDL}$ as n function of $V_{DG}$.

## V. RESULTS AND DISCUSSIONS

Our model is compared with both the experimental result and the models designed by previous researcher [4]. Gate induced drain leakage current is plotted against gate voltages for fixed drain voltage. It is shown that our model is fitted within three sigma error ranges of practical data which can be concluded as well matched model. Figure-8 emphasizes by magnifying various portions of Figure-7, that our model is well matched as it lied within acceptable error range of the experimental data Sample specifications are given below:

Sample Specifications: W=0.8 nm; $T_{ox}$=70 Å; L=20μm
$N_d$=5×10$^{19}$ gm/cc. $N_d$=1.549×10$^{-9}$ cm$^2$; $V_{FB}$=0.83eV; $\epsilon_{si}$= 3.9eV; $\epsilon_{ox}$= 12eV; T=300K and K = 1.38× 10$^{-23}$ is Boltzman Constant.
Model Parameter Specifications:
α=2.074×10$^{-26}$ and β=1.85×10$^{-18}$

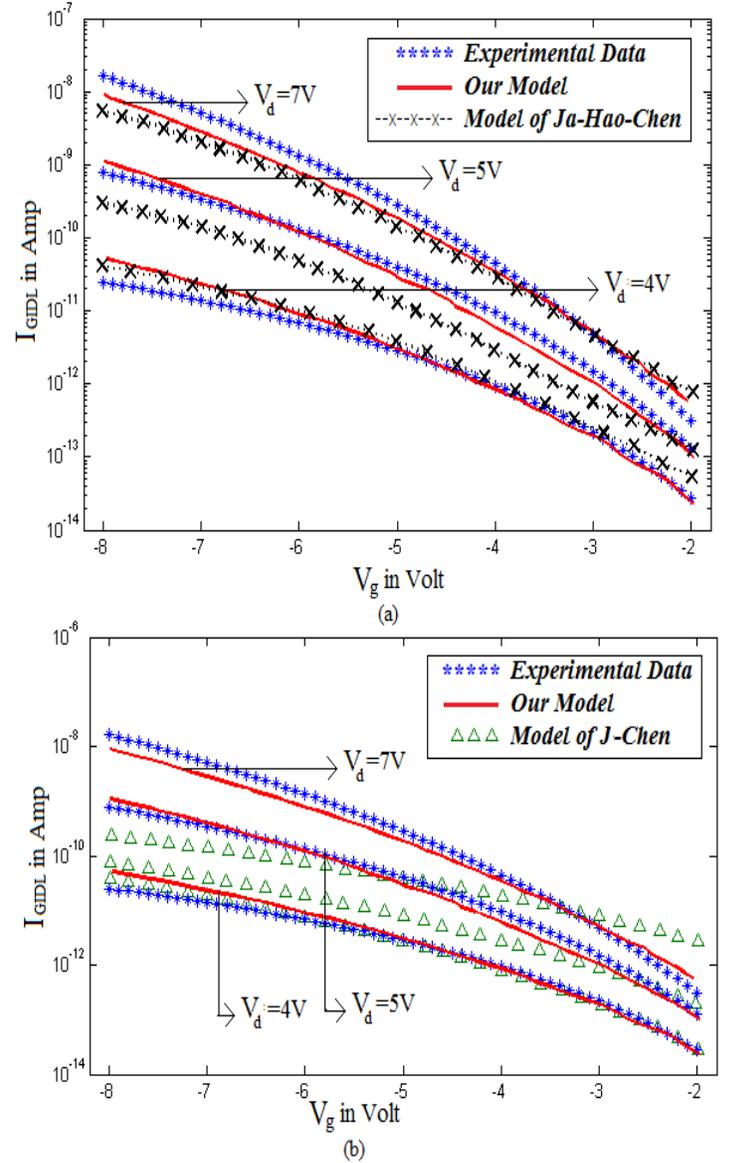

Figure-7: Model Validity by comparing with (a) Ja-Hao-Chen's Model[4] and (b) J-Chen's Model.[1]

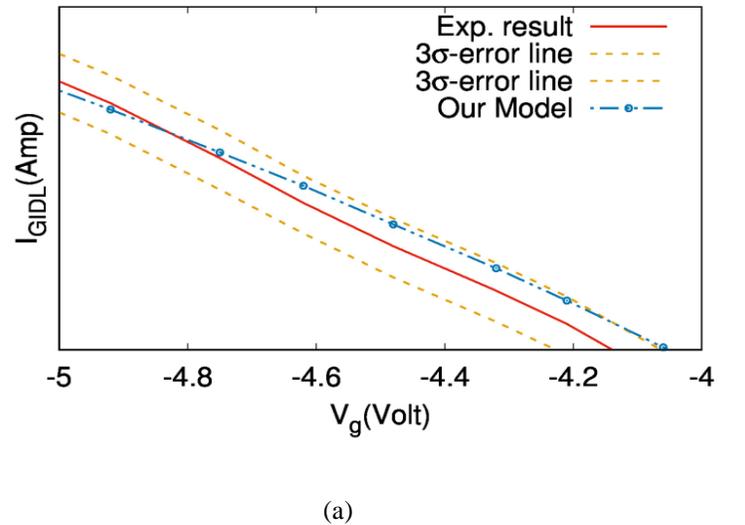

(a)

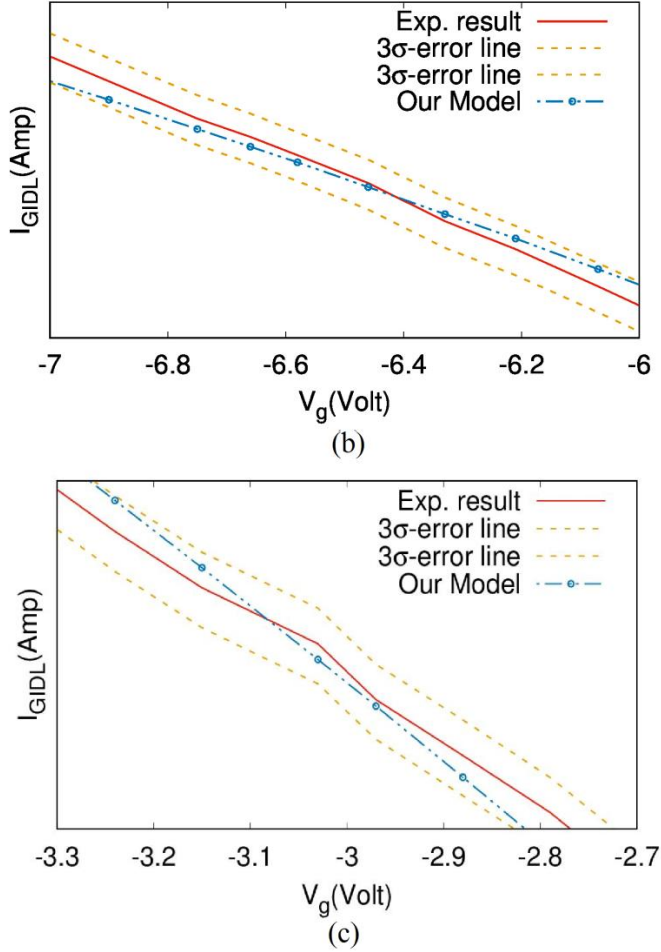

Figure-8: Error calculation with 3-sigma range ($\pm 11.83\%$) for (a) $V_d=4V$ (b) $V_d=5V$ and (c) $V_d=7V$.

## VI. Conclusion

An analytic expression of gate induced drain leakage current is proposed and verified with experimental data and other previous models. Results are well matched and well fitted within acceptable error range with experimental data. Bezier Curve approach is the key concept to establish our model so all calculations could be done in n easy manner and the extracted expression is too handy to frequent use.

## Appendix

### A. Appendix 1

Equation Formation from Bezier Curve Control Points:

$$P_0 \equiv [x_1, y_1], P_1 \equiv [x_2, y_2], P_3 \equiv [x_3, y_3]$$

Now, the equations formed by the control points are as follows

$$x = (1-t)^2 x_1 + 2(1-t)t x_2 + t^2 x_3$$
$$\text{And } y = t^2 y_3$$

After eliminating t and neglected the terms where $y_3$ appears in denominator we get,

$$y \cong \frac{(x-x_1)^2 y_3}{4(x_2-x_1)^2}$$

Here the Bezier Curves are used to model n band diagram. Thus obviously x and y axis denotes distance (in micro order) and energy level (in eV order) respectively. The term A is constant of proportionality and set in 400. Thus either $y_3$ or $Ay_3$ appears in denominator and the numerator contains x or x axis related terms the factor can be easily neglected.

### B. Appendix 2

Solving Integral of WKB Approximation:

$$y(x_n) = \int_a^b \left[ \frac{\left(x_a - \frac{1}{2}\right)^2 AV_b}{4\left(\frac{1}{2}-x_{n2}\right)\left\{1-\left(\frac{1}{2}+x_{n2}\right)\right\}} - E \right]^{\frac{1}{2}} dx_n$$

$$or, \quad y(x_n) = \frac{\sqrt{AV_b}}{2\left(\frac{1}{2}-x_{n2}\right)} \int_a^b \left\{ \left(x_n - \frac{1}{2}\right)^2 - \frac{4E\left(\frac{1}{2}-x_{n2}\right)^2}{AV_b} \right\}^{\frac{1}{2}} dx_n$$

Now the integral is in the form of $\int \sqrt{x^2-a^2} \, dx$ whose solution is $\frac{x\sqrt{x^2-a^2}}{2} - \frac{a^2}{2}\log\left|x+\sqrt{x^2-a^2}\right|$.

## Acknowledgment

This research was supported by Jadavpur University, Calcutta, India. We thank our colleagues from Dept. of Physics, Jadavpur University who provided insight and expertise that greatly assisted the research. At first, we would like to show our gratitude to Prof. D. P. Bhattacharyya for sharing his pearls of wisdom. We thank Ms. D. Basu, Senior Research Scholar, Dept. of Physics, Jadavpur University for assistance with her programming skill and Mr. B. Roy, Junior Research Fellow, Dept of Physics, Jadavpur University for contributing his magnificent model designing concept. In addition to that we are also immensely grateful to Mr. A. Chatterjee, Senior Research Fellow, Dept. of Physics, Ramakrishna Mission Vivekananda University for sharing his outstanding knowledge of data analysis. Moreover Mr. A. Das, M.Sc. Electronics from University of Calcutta, this name should be highlighted for his constant assistance and support.